# Text Analysis of ETDs in ProQuest Dissertations and Theses (PQDT) Global (2016-2018)

**Manika Lamba**
Research Scholar, Department of Library and Information Science, University of Delhi, Delhi-110007.( E) mlamba@libinfosci.du.ac.in

## Abstract

The information explosion in the form of ETDs poses the challenge of management and extraction of appropriate knowledge for decision making. Thus, the presentstudy forwards a solution to the above problem by applying topic mining and prediction modeling tools to 263 ETDs submitted to the PQDT Global database during 2016-18 in the field of library science. This study was divided into two phases. The first phase determined the core topics from the ETDs using Topic-Modeling-Tool (TMT), which was based on latent dirichlet allocation (LDA), whereas the second phase employed prediction analysis using RapidMinerplatform to annotate the future research articles on the basis of the modeled topics. The core topics (tags) for the studied period were found to be *book history*, *school librarian*, *public library*, *communicative ecology*, and *informatics* followed by text network and trend analysis on the high probability co-occurred words. Lastly, a prediction model using Support Vector Machine (SVM) classifier was created in order to accurately predict the placement of future ETDs going to be submitted to PQDT Global under the five modeled topics (*a* to *e*). The tested dataset against the trained data set for the predictive performed perfectly.

## Introduction

In addition to journal articles, Electronic Theses and Dissertations (ETDs) are the most frequent type of educational resource which is consulted by the scientific community from time to time. "The submission of theses and dissertations in electronic format has opened the door for the user community to have an entrance to the knowledge implanted in these works through different national and international ETDs and databases" (Haneefa and Divya, 2018). "They are well defined and well-referenced administrative documents on both the national and international level. They play an important role in research by adopting a Knowledge Organization System (KOS) architecture to enhance Information Retrieval (IR) systems and their performance" (Gunjal and Gaitanou). As the number of textual data including ETDs are increasing exponentially every day over the Web, the issue of organizing, managing and disseminating information has attracted attention and led to many efforts, including the Knowledge Management, Content Analysis, Text Analysis, Text Classification, Text Categorization, Search Strategy, Linked Data, and Semantic Web, etc. to enhance information retrieval systems and their performance for decision making.

"During the last decade, the growth of ETDs has been increased tremendously among universities and other organizations all over the world"(Gunjal and Gaitanou).The information explosion in the form of ETDs poses the challenge of management and extraction of appropriate knowledge for decision making. Thus, this study forwards a solution to the above problem by applying topic mining and prediction modeling tools to ETDs submitted in ProQuest Dissertations and Theses (PQDT) Globaldatabase during 2016-2018 in the field of library science.This present study (i) discovers the hidden topical pattern, (ii) performs text network and trend analysis for the highest frequency words generated by topic modeling; and (iii) presents a best fitted predictive model for the ETDs retrieved.

## Literature review

Some of the prominent studies existing in literature on the topic mining with respect to ETDs are by Lamba and Madhusudhan (2018), "applied topic modeling to Library and Information Science (LIS) theses submitted to Shodhganga (an Indian ETDs digital repository) to determine the five core topics/tags and then the performance of the built model based on those topics/tags were analysed". Sugimoto et al. (2011), "identified the changes in dominant topics in library and information science (LIS) over time, by analyzing the 3,121 doctoral dissertations completed between 1930 and 2009 at North American Library and Information Science programs. The authors utilized Latent Dirichlet Allocation (LDA) to identify latent topics diachronically and to identify representative dissertations of those topics". Morgan et al. (2008), used "OSCAR3, an Open Source chemistry text-mining tool, to parse and extract data from theses in PDF, and from theses in Office Open XML document format". Brook et al. (2014), emphasized on the fact that the "main barriers against the uptake of TDM were not technical, but, primarily a lack of awareness among the academics, and a skills gap. They further elaborated on the legal issues around the copyright, database rights, and to some policy choices of restrictions being implemented by publishers". Schöpfel et al. (2015), highlighted that "the legality of mining ETDs has to be ensured by a legal Text Data Mining exception; moreover, the issuing of prescription rules should systematize a third party agreement to clear rights in a mining context. Prescription rules could also ease the feasibility by proposing application standards and by promoting rich metadata and text structures". Nanni and Paci (2017), studied the "hermeneutic and text mining practices while analyzing one of the primary research output of European universities, namely doctoral theses and present an enriched dataset".

Some of the studies which apply prediction modeling are by Lamba and Madhusudhan (2019), who "described the importance and usage of metadata tagging and prediction modeling tools for researchers and librarians. 387 articles were downloaded from DESIDOC Journal of Library and Information Technology (DJLIT) for the period 2008–17. This study was divided into two phases. The first phase determined the core topics from the research articles using Topic-Modeling-Toolkit (TMT), which was based on latent dirichlet allocation (LDA), whereas the second phase employed prediction analysis using RapidMiner toolbox to annotate the future research articles on the basis of the modeled topics"; Özmutlu and Çavdur (2005), who "proposed an artificial neural network to identify automatically topic-changes in a user session by using the statistical characteristics of queries, such as time intervals and query reformulation patterns"; and Benton et al. (2016), who considered "survey prediction from social media. They used topic models to correlate social media messages with survey outcomes and to provide an interpretable representation of the data. Rather than rely on fully unsupervised topic models, they used existing aggregated survey data to inform the inferred topics, a class of topic model supervision referred to as collective supervision. They introduced and explored a variety of topic model variants and provided an empirical analysis, with conclusions of the most effective models for this task".

## Methodology

The method followed for this study wasa composite one as the study dealt with the three important aspects of the problem (i) information retrieval, (ii) the designing, and (iii) the evaluation with the help of topic mining (Topic-Modeling-Tool, 2011), text network and trend analysis (VoyantTools, 2019), and prediction modeling (RapidMiner, 2019). A total of 263 ETDs were retrieved in the English language for library science subject from PQDT Global dataset for 2016-18. Out of the 2563 ETDs,10 ETDs were found without advisor's name and 7 ETDs were found without department's name but those ETDs were still included in the study as the incomplete bibliographic details did not compensate the topic and prediction modeling process directly.The study wasdividedinto two phases. In the first phase, ETDs on library science subject for the period 2016-18 was downloaded from PQDT Global database and converted to text format (information retrieval), followed by the analysis of the text corpus according to latent dirichlet allocation (LDA) probabilistic topic modeling method with the help of Topic-Modeling-Tool (designing and evaluation). Five topics were identified for the studied period and each topic contained a probability value. The topics were then ranked by probability values and only the top five representative topics were selected. Similarly, the probability of each word was calculated to represent the association between a word and the given topic and the top five words werechosen as most representative of the topic. Further, text network and trendanalysis were made using the 25 high-probability co-occurrencewords for all the five topics to have a better insight into the data (evaluation). In the second phase, prediction analysis was performed with the help of a text mining platform called RapidMiner. The process included the following steps (designing and evaluation):

➢ Pre-processing of the documents (i.e. tokenization, stemming, filtering stop-words, transforming the cases, and generating n-grams per terms);

- ➢ Splitting the data into two subsets;
- ➢ Training and testing of the data using split validation;
- ➢ Application of the appropriate classifier to build the predictive model; and
- ➢ Measuring the performance of the model.

**Latent Dirichlet Allocation (LDA)**

"This paper focuses on the use of LDA (Blei, Ng, and Jordan 2003), which is based on Dirichlet distribution to model the topics from the corpus of LIS ETDs. In this study, each ETD got represented as a pattern of LDA topics making every ETDappear. LDA automatically inferred the topic discussed in a collection of ETDs and these topics could be used to summarize and organize ETDs. LDA is based on probabilistic modeling and the observed variables are the bags of words per ETD whereas hidden random variables are the topic distribution per ETD" (Lamba and Madhusudhan, 2018). "The main goal of LDA is to compute the posterior of the hidden variables given the value of the observed variables" (Allahyari et al. 2017). "The assumptions of LDA for the study were: (i) ETDs with similar topics would use similar groups of words, (ii) ETDs were a probability distribution over latent topics, and (iii) topics were probability distributions over words" (Lamba and Madhusudhan, 2018).

# Results

**Topic Analysis**

On the basis of the output files (present in both CSV and HTML formats) generated by the Topic-Modeling-Tool (TMT), a comprehensive analysis had been performed for the studied period.After topic modeling had been conducted to the full-text corpus of the ETDs extracted from PQDT Global databaseusing the TMT, analysis of the output files had been undertaken to generate knowledge and to assign appropriate *topics* to the group of words generated. Table 1 summarizes the LDA result for the ETDs. "It showed the labeling of the topics, *a* through *e*, which were organized in descending order according to their probability values (where *a* having the highest probability value). It summarized the core topics which might be considered as the hot research trend for the corresponding period. It further listed the word co-occurrence pattern over time and summarized the top five words or the high loading keywords, ranked by the probability value for each period in the descending order. Thus, topic analysis is the process of assigning *topics* to a group of higher frequency words arranged in decreasing order and analyzing the results generated from the automated tool for the purpose of management and organization of the text documents" (Lamba and Madhusudhan, 2019). Further, in addition to the groups of words, representative ETDs were also consulted simultaneously to label the *topics* appropriately. Representative ETDswere the five core ETDs ranked on the basis of the highest topic proportion percentage for the given modeled topic (Table 2).

Five topics were modeled for the studied period (where number of ETDs=263; number of Topics=5; α=10.0; β=0.01) where the evidence from high-loading keywordsand most representative ETDs showed that *Topic a* was about *book history*with an emphasis on americanhistory and library whereas *Topic b* was about *school librarian* with a focus on students, information, and research. *Topic c* was on *public library* with a focus on community. Representative ETDs and keywords for *Topic d* indicated a focus on *communicative ecology* with an emphasis on social, health, media, and family. Lastly, *Topic e* was on *informatics* with a focus on data, research, knowledge, and search.

**Table 1.** Latent dirichlet allocation results for the period 2016–18 (263 ETDs)

| *Topic a* **Bookhistory** | **Topic b School librarian** | **Topic c Public library** | **Topic d Communicative ecology** | **Topic e Informatics** |
|---|---|---|---|---|
| library | students | library | social | information |
| books | school | libraries | health | data |
| history | information | public | information | research |
| book | research | community | media | knowledge |
| american | librarians | study | family | search |

**Table 2.** Titles corresponding to the representative ETDs for 2016-18 (263 ETDs)

|  | Topic a | Topic b | Topic c | Topic d | Topic e |
|---|---|---|---|---|---|
| Representative Title 1 | Exploring racial diversity in Caldecott Medal-winning and honor books | Factors that Influence Middle School Mathematics Teachers' Willingness to Collaborate with School Librarians | What Happens When Entrepreneurial Public Libraries Change Directors? | School Librarians' Perception of Adopting E-books in their School MediaCenters: A Multiple-case Study | Bootstrap-Based Confidence Intervals in Partially Accelerated Life Testing |
| Representative Title 2 | Books about music in Renaissance print culture: Authors, printers, and readers | Examining Middle School Teacher Practices and Attitudes Regarding Teaching Information Literacy Skills | Older Voluntarism and Rural Community Sustainability: A Case Study of a Volunteer-based Rural Library | Interactions in Calls to the 9-1-1 Emergency System in Costa Rica | Scripts in a frame: A framework for archiving deferred representations |
| Representative Title 3 | Judging a Book by Its Cover: The Context Book Covers Provide | Female Saudi Pre-Service Teachers' Competency in Information Literacy, Perceptions of Future | After-School Activities Policy and the Atlanta Fulton Public Library System | Information Practices Relative to Parental Mediation and the Family Context | Representing the Search Session Process |

| | | Classroom Practice, and the Role of Librarians | | Among Puerto Rican and Dominican Teens | |
|---|---|---|---|---|---|
| Representative Title 4 | Class Acts: The Twenty-Fifth and Twenty-sixth Earls of Crawford and Their Manuscript Collections | From Information Experts to Expert Educators? Academic Librarians' Experiences with Perspective Transformation and their Teaching Identities | Organizational Culture and Library Chief Executive Officers' Servant Leadership Practices | No End in Sight: A Critical Discourse Analysis of U.S. National Newspaper Coverage of the Iraq War | Research and innovation in West Africa: An informetric analysis within the framework of the Triple Helix model |
| Representative Title 5 | Exploring the Convenience Versus Necessity Debate Regarding SCI-HUB Use in the United States | Academic Librarians' Teacher Identity Development through the Scholarship of Teaching and Learning: A Mixed Methods Study | Diversifying Funds to Enhance Financial Sustainability of a County Library System | Distant close ties: Jamaican immigrants, mediated communication, and the primacy of voice | H3DNET: A Deep Learning Framework for Hierarchical 3D Object Classification |

*Word Analysis*

The co-word pattern generated from TMT was further analyzed using text network and trend analysis to get a better insight into the hidden word pattern. As the data was too big for VoyantTools to process, only bibliographic data including, abstract, title, author, advisor, keywords, subject, etc. was used as the corpus in the VoyantTools. Figure-I shows the trend-line graph for the corpus. "Trend-line graph depicts the distribution of a word's occurrence across a corpus. It is a visualization that represents the frequencies of terms across documents in a corpus or across segments in a document, depending on the mode. The relative frequency determines the term frequency in a document whereas raw frequency is the absolute count for each document" (VoyantTools, 2019). The asterisk (*) shows the search syntax to trigger a search for the match terms as one term, for instance, 'coat' will match the exact term 'coat' whereas 'coat*' will match terms that start with 'coat', 'coating', 'coats' etc. as one term. Further,"the table view showed the following columns:

➢ Term: this is the document term
➢ Count: this is the raw frequency of the term in the document
➢ Relative: this is the relative frequency (per 10 million words) of the term in the document
➢ Trends: this is a sparkling graph that shows the distribution of the term within the segments of the documents" (VoyantTools, 2019)

For this study, 19 high-probability co-occurrence words produced by TMT were queried in the search bar to determine the trend-line graph for the corpus. It can be observed from Figure 1 that the word *library*had the highest count and relative values whereas *family\** had the lowest. Further, the top 5 words with the highest count and relative values in the corpus were*library, information\*, study\*, research\*,* and *school\** in comparison to the words *healt\*, search\*, american\*, history,* and*family\** which had the lowest count and relative values.

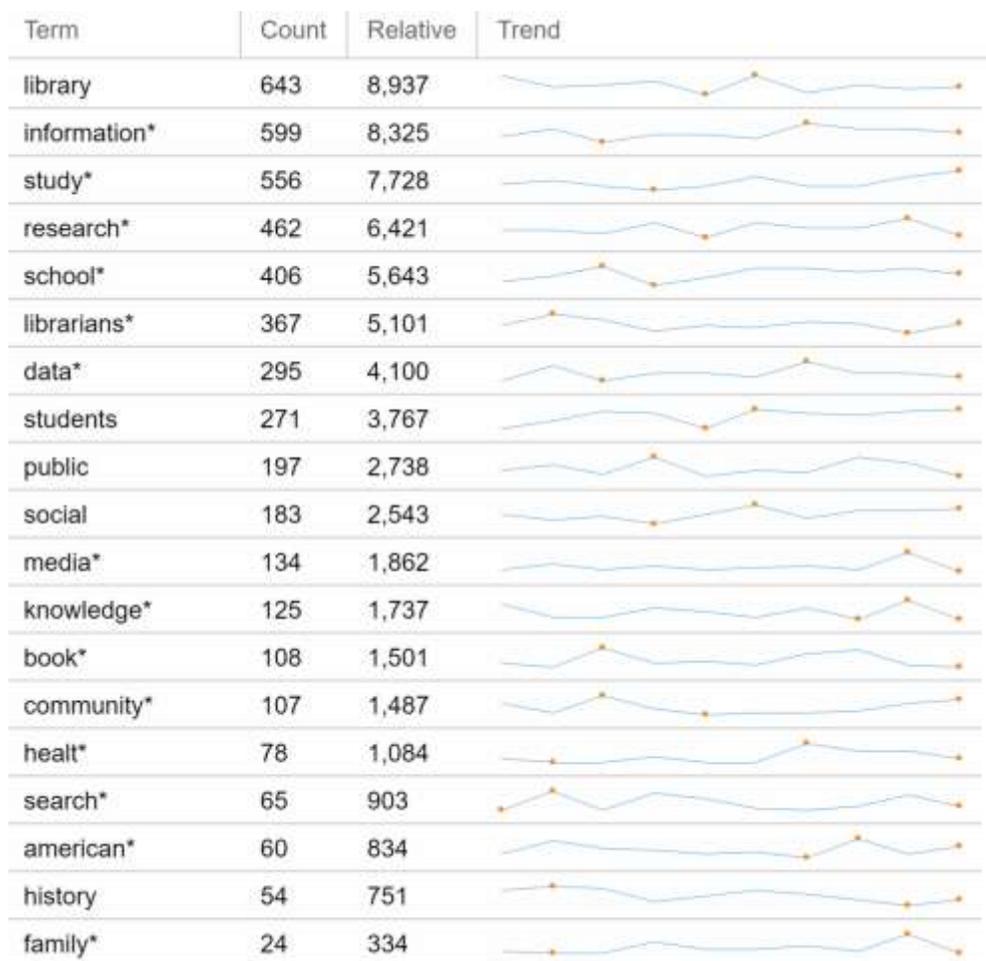

**Figure 1.** Trend-Line Graph for Co-Word Occurrence using VoyantTools

Figure 2 showed a "*collocate graph* which represents keywords and terms that occur in close proximity as a force-directed network graph" (VoyantTools,2019). "The *context slider*inVoyantTools determines how many terms to include for collocation. The value specifies the number of words to consider on each side of the keyword" (VoyantTools,2019). For this study, the context slider was set to its default value of 5 words per query.To make the text network graph for each modeledtopic, all the high probability words were added to the query box for the corpus to make the text network graph for the respective modeled topic. Further, the words were centralized for better insight. Further, Table 3 was prepared using Figure-II to determine the associated terms related to the co-occurred words in the text network graph for the respective modeled topics.

**Table 3.**Associated Terms Related to Co-Occurred Words in Text Network Graph

| **Modeled Topics** | **High Probability Co-occurred Words(Counts)** | **Associated Terms in Text Network Graph** |
|---|---|---|
| Topic-a | *library(643)* | school, book, history, american, information, study, science |
| | *books(61)* | collections, digital, librarians, school, music |
| | *history(4)* | public, era, library, book |
| | *book(46)* | history, library, club/s |
| | *american (56)* | male, athletes, association, information, and library |
| Topic-b | *students(271)* | information, research, study, library |
| | *school(361)* | library/libraries, librarian/s |
| | *information(33)* | seeking, science, literacy, library, research |
| | *research(16)* | libraries, study, information, question/s |
| | *librarians(128)* | library, study, faculty, teachers |
| Topic-c | *library(643)* | study, information, public, libraries, school, science, community |
| | *libraries(253)* | library, associations, public, information, college, study |
| | *public(197)* | school, library/libraries, study |
| | *community(106)* | members, based, college, information, library |
| | *study(31)* | library, libraries, public, information, mixed, research, findings |
| Topic-d | *social(183)* | capital, justice, data, media |
| | *health(68)* | insurance, seeking, information, family |
| | *information(20)* | seeking, health, science, library, literacy |
| | *media(34)* | static, school, specialists, non, use |
| | *family(23)* | health, topics, related, history, behaviors |
| Topic-e | *information(594)* | search, library, seeking, science, literacy, research, knowledge |
| | *data(273)* | using, collection, analysis, collected |
| | *research(406)* | question/s, study, libraries, information |
| | *knowledge(121)* | based, domain, organization, library, information |
| | *search(47)* | information, mediation, survey, process, social |

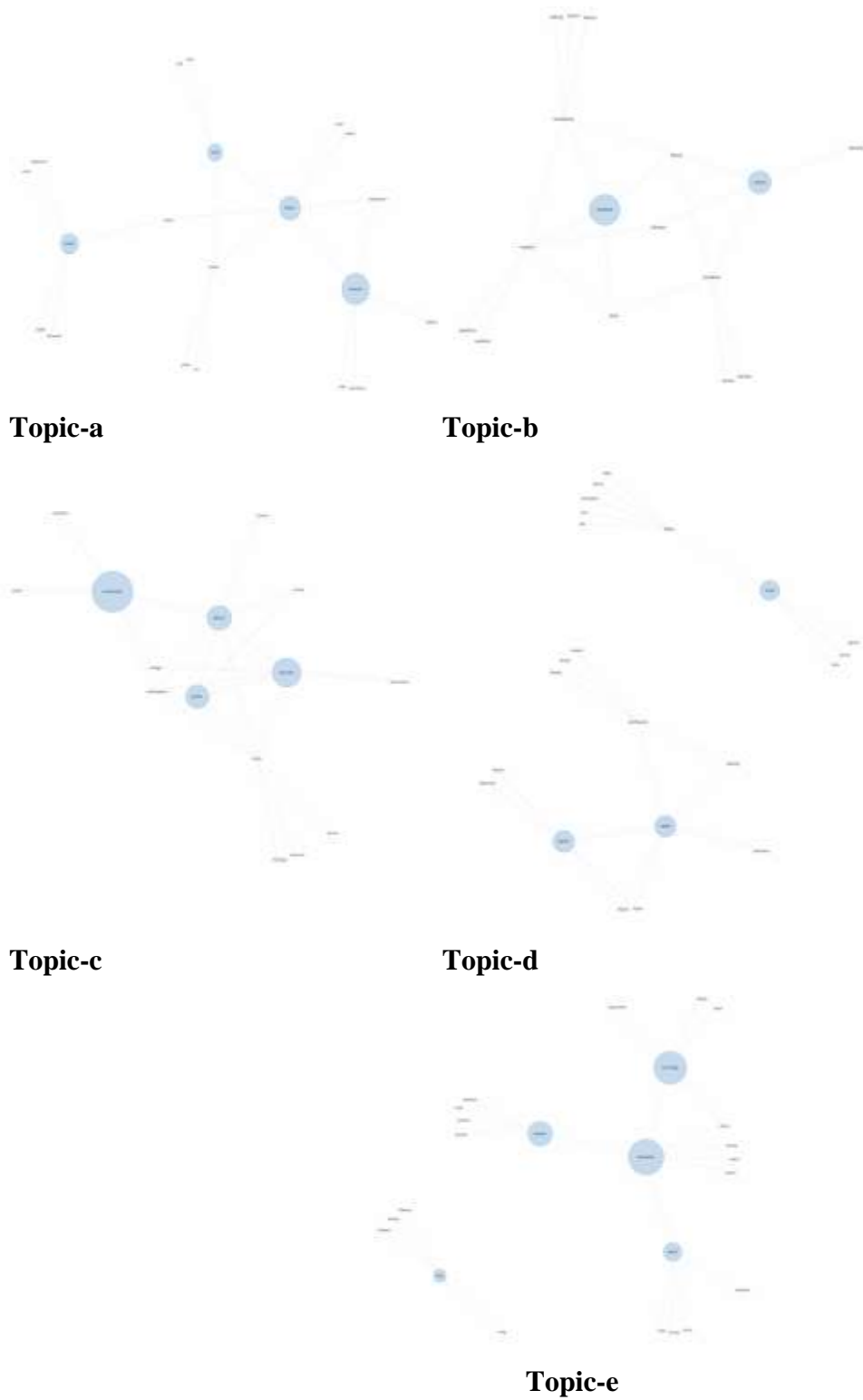

**Topic-a**  **Topic-b**

**Topic-c**  **Topic-d**

**Topic-e**

**Figure 2.** Text Network Graph of Co-Word Occurrence using VoyantTools

**Prediction Modeling**

Support Vector Machine (SVM) classifier was used to create the prediction model. The model was created using 263 tagged ETDs (*a* to *e*), where 70% (184) of the data was allocated to the training set and 30% (79) was allocated to test set randomly using the split validation technique.Once the parameters of the models were finalized, the testing set was run through the model. The actual test class was compared to the predicted class to determine the kappa, precision, and recall values. Figures-III showsperfect values for the tested data set against the trained data set for the predictive model.

| kappa: 1.000 | true Topic a | true Topic b | true Topic c | true Topic d | true Topic e | class precision |
|---|---|---|---|---|---|---|
| pred. Topic a | 41 | 0 | 0 | 0 | 0 | 100.00% |
| pred. Topic b | 0 | 91 | 0 | 0 | 0 | 100.00% |
| pred. Topic c | 0 | 0 | 57 | 0 | 0 | 100.00% |
| pred. Topic d | 0 | 0 | 0 | 20 | 0 | 100.00% |
| pred. Topic e | 0 | 0 | 0 | 0 | 54 | 100.00% |
| class recall | 100.00% | 100.00% | 100.00% | 100.00% | 100.00% | |

**Figure 3.** Screenshot of Evaluation of Prediction Analysis using RapidMiner

## Conclusion

The study used three different tools to perform topic modeling, text network analysis, trends analysis, and prediction modeling. Topic modeling was performed to tag the corpus of full-text LIS ETDs submitted to PQDT Global for the epoch 2016-18. The core topics (tags) for the studied period were found to be *book history*, *school librarian*, *public library*, *communicative ecology*, and *informatics*. LIS ETDs on the PQDT Global website can be tagged with the modeled topicsto have a faster information retrieval searching experience by the users. The limitations of the study include the prior identification of an appropriate number of topics for the ETDs before performing LDA; the incompetence of the Dirichletalgorithm to correlate among topics; and lastly, the manual interpretation of labeling of topics.The present study then applied text network and trend analysis on the high probability co-occurred words to have a better insight into the results.Further, a prediction model using Support Vector Machine (SVM) classifier was createdin order to accurately predict the placement of future ETDs going to be submitted to PQDT Global under the five modeled topics (*a* to *e*).The tested dataset against the trained data set for the predictive performed perfectly. The limitation of using prediction modeling for the study was that the dataset was not truly representative of LIS ETDs of the database. The training of the model to learn and fit the parameters could be done perfectly if more data is taken into account. This work will have a broad application to those interested in information retrieval of ETDs. The findings of the study will help the users in faster information retrieval from PQDT Global database by searching the ETDs on the basis of the concept/theme behind each ETD instead of subject, title, keywords, author, year of completion of the ETD, advisor, university, department, etc.

ICDL 2019: Research Data Management# References

1. Allahyari, M., Pouriyeh, S., Assefi, M., Safaei, S., Trippe, E. D., Gutierrez, J. B., and Kochut, K. 2017.**A Brief Survey of Text Mining: Classification, Clustering and Extraction Techniques**. *ArXiv:1707.02919 [Cs]*. http://arxiv.org/abs/1707.02919 (accessed on 21st July 2019).

2. Benton, A., Paul, M. J., Hancock, B., and Dredze, M. 2016. **Collective Supervision of Topic Models for Predicting Surveys with Social Media**.*InProceedings of Thirtieth AAAI Conference on Artificial Intelligence (AAAI)*, 2892–2898.

3. Blei, D. M., Ng, A. Y., and Jordan, M. I. 2003.**Latent dirichletallocation**.*Journal of Machine Learning Research* Volume no 3(1):993–1022.

4. Brook, M., Murray-Rust, P., and Oppenheim, C. 2014.**The Social, Political and Legal Aspects of Text and Data Mining (TDM)**.*D-Lib Magazine*, Volume no 20(11/12). https://doi.org/10.1045/november14-brook(accessed on 21st July 2019).

5. Divya, P. and Haneefa, M. 2018. **Digital Reading Competency of Students: A Study in Universities in Kerala**. *DESIDOC Journal of Library & Information Technology* Volume no 38(2):88–94. https://doi.org/10.14429/djlit.38.2.12233

6. Gunjal, B., and Gaitanou, P. 2015. **ETDs and Open Access for Research and Development: Issues and challenges**. *In18th International Symposium on Electronic Theses and Dissertations Evolving Genre of ETDs for Knowledge Discovery,* Delhi, India, organised by Jawaharlal Nehru University.https://doi.org/10.13140/rg.2.1.2826.4401

7. Lamba, M. and Madhusudhan, M. 2018**.**Metadata Tagging of Library and Information Science Theses: Shodhganga (2013-2017)**.***In ETD 2018 Taiwan Beyond the Boundaries of Rims and Oceans: Globalizing Knowledge with ETDs,* Taipei, Taiwan, organised by National Central Library. https://etd2018.ncl.edu.tw/images/phocadownload/3-2_Manika_Lamba_Extended_Abstract_ETD_2018.pdf (accessed on 21st July 2019).

8. Lamba, M. and Madhusudhan, M. 2019.Mapping of topics in DESIDOC Journal of Library and Information Technology, India: a study.*Scientometrics.*https://link.springer.com/article/10.1007/s11192-019-03137-5 (accessed on 21st July 2019).

9. Lamba, M. and Madhusudhan, M. 2019.**Metadata Tagging and Prediction Modeling: Case Study of DESIDOC Journal of Library and Information Technology (2008-2017).** *World Digital Libraries: An International Journal* Volume no 12(1):33-89.

10. Morgan, P., Downing, J., Murray-Rust, P., Stewart, D., Tonge, A., Townsend, J. A., and Rzepa, H. S. 2008.**Extracting and re-using research data from chemistry e-theses: The SPECTRa-T project.**http://www.dspace.cam.ac.uk/handle/1810/230116(accessed on 21st July 2019).

11. Nanni, F., and Paci, G. 2017.**A Discipline-Enriched Dataset for Tracking the Computational Turn of European Universities.***In Proceedings of the 6th International Workshop on Mining Scientific Publications*, pp. 29–33.

12. Özmutlu, S., and Çavdur, F. 2005.Neural network applications for automatic new topic identification.*Online Information Review*Volume no 25(1):34–53. https://doi.org/10.1108/14684520510583936
742